\documentclass{iau}

\usepackage{graphicx}
\usepackage{multirow}
\usepackage{txfonts}
\usepackage{aas_macros}
\begin{document}

\lefttitle{Cambridge Author}
\righttitle{Proceedings of the International Astronomical Union: \LaTeX\ Guidelines for~authors}

\jnlPage{1}{7}
\jnlDoiYr{2025}
\doival{10.1017/xxxxx}
\volno{395}
\pubYr{2025}
\journaltitle{Stellar populations in the Milky Way and beyond}

\aopheadtitle{Proceedings of the IAU Symposium}
\editors{J. Mel\'endez,  C. Chiappini, R. Schiavon \& M. Trevisan, eds.}

\title{Chronology of our Galaxy from Gaia CMD-fitting (ChronoGal): the early formation of the Milky Way disk and the impact of Gaia-Sausage-Enceladus}

\author{Carme Gallart$^{1,2}$, Emma Fernández-Alvar$^{1,2}$, Santi Cassisi$^{3,4}$, \\Tomás Ruiz-Lara$^{5,6}$, Francisco Surot$^1$, Guillem Aznar-Menargues$^2$, \\Yllari González-Koda$^5$, David Mirabal$^2$, Anna B. Queiroz$^{1,2}$, Alicia Rivero$^{1,2}$}
\affiliation{(1) Instituto de Astrofísica de Canarias, Tenerife, Spain; (2) Universidad de La Laguna, Tenerife, Spain; (3) INAF -- Astronomical Observatory of Abruzzo, Teramo, Italy; (4) INFN, Sezione di Pisa, Pisa, Italy ; (5) Universidad de Granada, Departamento de F\'isica Te\'orica y del Cosmos, Granada, Spain; (6) Instituto Carlos I de F\'isica Te\'orica y computacional, Universidad de Granada, Granada, Spain. }

\begin{abstract}
The derivation of precise stellar ages is considered the current major challenge to reconstruct the chronology of the Milky Way. Color-magnitude diagram (CMD)-fitting offers a robust alternative to individual age determinations via the derivation of dynamically evolved star formation histories (deSFH) and age-metallicity distributions \citep{Gallart2024}. Our new suite of routines, CMDft.Gaia, specifically developed to analyse Gaia CMDs, produce deSFHs which are robust against sensible changes in the input parameters and extremely precise, providing an unprecedentedly detailed characterization of the successive events of star formation that, since its early evolution, have shaped the current Milky Way. Also important is the fact that, thanks to the high completeness of the Gaia photometric data, CMDft.Gaia provides the actual number of stars and the mass involved in the different events of star formation. 

The current analysis of the deSFH for stellar populations within 100 pc of the Sun, as well as for kinematically selected stars in the thin disk, thick disk, and halo, allows us to sketch a tentative picture of Milky Way evolution. The findings indicate that star formation commenced very early in a thick disk, with a small fraction of stars having [M/H]$\lesssim$-0.5 forming more than 12 Gyr ago. This phase culminated in a more prominent 12 Gyr old population with [M/H]$\simeq$-0.5. Approximately 11 Gyr ago, the merger with GSE triggered an intense burst of star formation, generating most of the thick disk’s mass and enriching its metallicity to solar levels. Subsequently, the bulk of the star formation in the thin disk started and continues with a somewhat episodic behaviour up to the present time. 

\end{abstract}


\maketitle
 
\section{Introduction}
The last few years have seen an enormous progress in Galactic Archaeology thanks to the large, high-quality datasets from the ESA mission Gaia \citep{GaiaDR3_2023Vallenari} and from a large number of ambitious ground-based spectroscopic surveys such as APOGEE \citep{2020ApJS..249....3A}, GALAH \citep{Buder2021_Galah} or LAMOST \citep{2020ApJS..251...27W}. The main goal of all these projects is to study the formation and evolution of the Milky Way, and in particular, to determine the star formation history (SFH) of its various components. Thanks to these data, we have now, for example, a detailed mapping of the chemical abundances across the Milky Way disk, which reveal the spectacular spatial variation of the high-$\alpha$ and low-$\alpha$ stellar populations over a wide range of galactocentric radius and height from the plane \citep{Queiroz2020_alfas}. Similarly, the local Galactic halo is now thought to be mainly composed by the debris of the Gaia-Sausage-Enceladus (GSE) dwarf galaxy \citep[considered to be the last major merger experienced by the Milky Way,][]{belokurov2018_sausage, helmi2018} as well as heated early disk stars \citep{Gallart2019Gaia, Belokurov2020Splash}, together with the debris of many other less massive mergers \citep[e.g.][]{Naidu2020, Dodd2023}. However, the current biggest challenge for a real breakthrough in the field is the difficulty to determine precise stellar ages for unbiased and large samples of stars \citep{Miglio2017_Plato}. 

Ages are indeed essential to establish a temporal sequence of events and to correctly interpret the measured properties in the context of galaxy formation and evolution models. {\it Precise ages} for sufficiently large samples of stars sampling the different Galactic components would allow us to answer fundamental questions about Milky Way's evolution such as: {\it When} did the dominant stellar disk of the Milky Way emerge? Is there an {\it early}, very metal poor thin disk? What is the {\it age}-metallicity distribution across the disk? {\it How long} did the high-$\alpha$ phase last? Did the high-$\alpha$ and low-$\alpha$ phases {\it overlap in time}? What is the role of mergers in the Milky Way disk formation and evolution?

In this contribution, we will discuss color-magnitude diagram (CMD) fitting as a key tool to provide the age dimension in Galactic Archaeology. This technique, which was previously limited to studying external Local Group galaxies, can now be applied to our Galaxy thanks to the availability of accurate Gaia-based distances. As a result, CMD-fitting is providing unprecedentedly precise and homogeneous age-metallicity distributions for millions of stars across the Milky Way. We will present here the study of several stellar samples in the disk and halo, as an example of the extraordinary potential of CMD-fitting in Galactic Archaeology. 

\section{Age determination in Galactic Archaeology: the role of CMD-fitting}

The most common approach to obtain temporal information in Galactic Archaeology is through the derivation of individual ages for stars with physical parameters derived from spectra, asteroseismology, or images, which are compared to a set of stellar evolution models which provide the age as a function of these parameters \citep{haywood2013, MiglioChiappini2021_Kepler, Sahlholdt2022MNRAS.510.4669S, Xiang_Rix_2022Natur.603..599X, Queiroz2023_starhorse, Pinsonneault2015_apokasc3}. ChronoGal \citep{Gallart2024}, instead, is using CMD-fitting to derive the Milky Way SFH from Gaia CMDs. In this approach, a best fit SFH is obtained by comparing the distribution of stars in the observed CMD with that in a large number of model CMDs, constructed from combinations of simple stellar populations, until the best match is found. Because the Milky Way is a very complex chemodynamical system, and a fraction of stars currently in a given volume are not necessarily born in it, we call dynamically-evolved (de)SFHs to the amount of mass transformed into stars, as a function of time and metallicity, somewhere in the Galaxy, to account for the population of stars currently found in a particular Milky Way volume.

\begin{figure}[h]
  \centering
  \includegraphics[scale=.47]{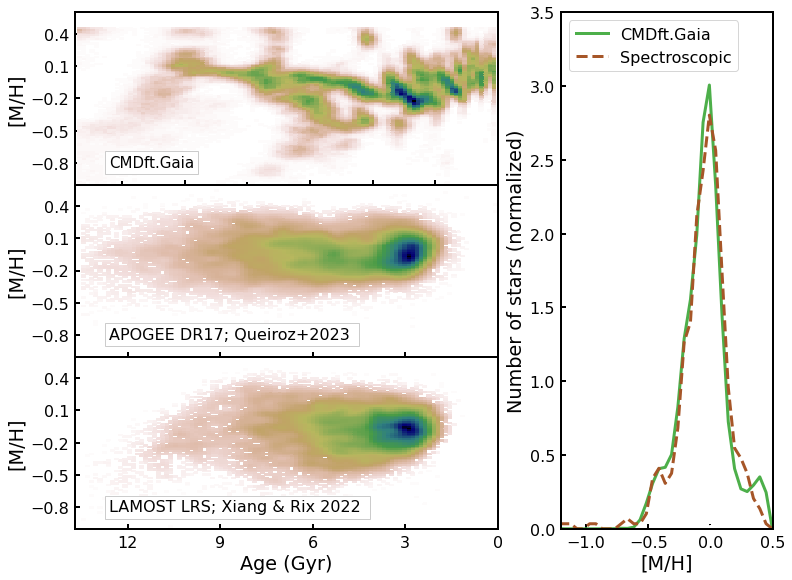}
  \caption{Comparisons of ChronoGal results with literature data. Left panel: Age-metallicity distributions derived with CMDft.Gaia for a cylinder of 250 pc radius and $|Z| \leq$ 500 pc, centred on the Sun \citep[ChronoGal’s approach,][top panel]{Gallart2024}, compared to those from individual star ages for a similar Milky Way volume \citep[][middle and bottom panels]{ Queiroz2023_starhorse, Xiang_Rix_2022Natur.603..599X}. Right panel: comparison of the ChronoGal metallicity distribution within 100 pc (green solid line) to that from \citet{Fuhrmann2017_LocalStarPops} spectroscopic sample (orange dashed line). }
  \label{compareMDF_singleStarsAMD}
\end{figure}

In \citet{Gallart2024} we have introduced CMDft.Gaia, a package specifically designed to derive SFHs through CMD-fitting with Gaia data. It includes ChronoSynth to perform the calculation of the synthetic {\it mother} CMDs, from which the simple stellar populations are extracted, DisPar-Gaia, the module in charge of the simulation of observational errors and completeness \citep[see][]{Fernandez-AlvarThinThick, Ruizlara2022_HS}, and DirSFH, the actual CMD-fitting routine. Through a comprehensive set of tests with mock stellar populations, \citet{Gallart2024} showed that {\it CMDft.Gaia produces deSFHs which are robust} against sensible changes in the input parameters (such as for example, the characteristics of the binary stars population, the stellar models or the number of stars in the mother CMD), {\it precise} (better than 5-10\% in age and 0.1 dex in metallicity in the dating of simple stellar populations), and {\it accurate} (with ages systematically overestimated by only a maximum of 6\% at intermediate-old ages).

From the deSFH, the age-metallicity distribution of the population under study can be obtained. Marginalizing over age, the metallicity distribution function can be obtained, which can be compared with metallicity distributions derived spectroscopically for the same population. The latter provides an excellent external check of the reliability of the method, as no information or assumptions on the chemical evolution or the metallicity distribution have been made in the deSFH derivation. The right panel of Figure \ref{compareMDF_singleStarsAMD} shows a comparison of the metallicity distribution derived from the deSFH of the Gaia Catalogue of Nearby Stars \citep[][the stellar sample within 100 pc of the Sun]{gcns} derived with CMDft.Gaia \citep[][solid green line]{Gallart2024} to that from the \citet{Fuhrmann2017_LocalStarPops} complete spectroscopic sample within 25 pc (dashed orange line). The impressive agreement between the two distributions is a proof of the reliability of the CMD-fitting results, for which only Gaia photometry and parallaxes have been used as input.

The left panels of Figure \ref{compareMDF_singleStarsAMD} provide a comparison of the ChronoGal age-metallicity distribution for a cylinder of 250 pc radius and $|Z| \leq$ 500 pc, centred on the Sun \citep[top panel, see][and Fernández-Alvar et al. in this conference]{Fernandez-AlvarThinThick}, with two age-metallicity distributions derived from state-of-the-art individual stellar ages catalogues \citet[][middle panel, based on APOGEE+Gaia]{Queiroz2023_starhorse} and from \citet[][bottom panel, based on LAMOST+Gaia]{Xiang_Rix_2022Natur.603..599X}. In those catalogues, only stars with guiding radius $7\le R_g\le 9$ and Z${_{\rm max}}\le 500$ pc \citep[according to the orbital parameters from][]{Kordopatis2023GaiaAges}, similar to the majority of the stars in the ChronoGal sample, are considered. These age-metallicity distributions from individual stellar ages provide much less detail, owing to larger age errors, than the one derived with CMD-fitting. Another important difference, not visually apparent, is that ChronoGal provides the actual number of stars and the mass involved in the different events of star formation. However, it does not provide individual stellar ages.

\section{The deSFH of the stars within 100 pc of the Sun}

\begin{figure}[t]
   \centering
  \includegraphics[scale=0.34]{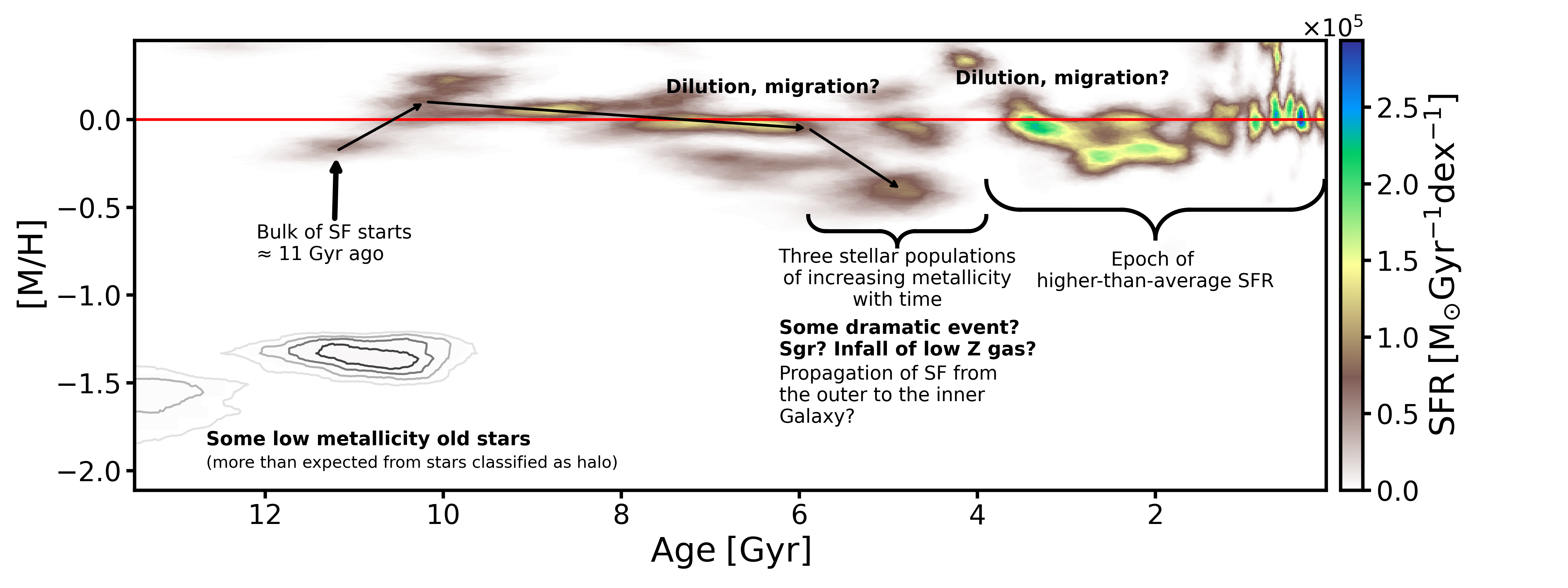}
  \caption{Annotated deSFH of the solar neighbourhood within 100 pc of the Sun. The main periods of star formation are indicated, together with hypothesis on the  possible origin of the various features.}
  \label{deSFH-gcns}
\end{figure}

In \citet{Gallart2024}, we derived the deSFH from the stars in the Gaia Catalogue of Nearby Stars \citep{gcns}, which contains the Milky Way stellar sample located within 100 pc of the Sun.  Figure \ref{deSFH-gcns}  shows the age-metallicity distribution of the mass transformed into stars as a function of lookback time (with old ages on the left) and metallicity, [M/H]\footnote{In our deSFH results, [M/H] refers to total metallicity, unlike some spectroscopic surveys in which the same [M/H] notation is used to refer to the iron abundance.}, obtained adopting the BaSTI-IAC \citep{Hidalgo2018basti} stellar evolution library, a \citet{kroupa1993} initial mass function, and a binary fraction $\beta=0.3$ with a minimum mass ratio $q_{\rm min}=0.1$ \citep[see][for details]{Gallart2024}. The colour bar indicates the star formation rate in units of M$_\odot$\,Gyr$^{-1}$\,dex$^{-1}$. This deSFH provides an unprecedentedly detailed view of the evolution of the Milky Way disk at the solar radius. The bulk of the star formation started between 11-10.5 Gyr ago with metallicity around solar (indicating a rapid chemical enrichment in the first couple of Gyr of evolution of our Galaxy) and continued with a somewhat decreasing metallicity trend until 6 Gyr ago, possibly the result of steady star formation fuelled by gas accretion while the radius of the Milky Way was progressively growing. In this scenario, the combined effects of gas dilution and chemical enrichment by ongoing star formation may be the origin of the observed trend \citep{Weinberg2017}. Between 6 and 4 Gyr ago, a notable break in the age-metallicity distribution is observed, with three stellar populations with distinct metallicities (sub-solar, solar, and super-solar), possibly indicating a dramatic event in the life of our Galaxy, such as the first infall of the Sagittarius dwarf galaxy, thought to have occurred around this epoch \citep{law2010, Ruiz-Lara2020Sgr}. Star formation then resumed 4 Gyr ago with a somewhat episodic behaviour, near-solar metallicity and average star formation rate higher than in the period before 6 Gyr ago. Interestingly, our results reveal the presence of intermediate-age populations exhibiting both a metallicity typical of the thick disk, approximately ${\rm[M/H]}\simeq-0.5$, and super-solar metallicity, likely indicating stellar migration from the outer and inner disk, respectively. Finally, a minor old, low metallicity population is also observed in this deSFH, implying a larger number of stars than those classified as halo in this volume. Therefore, we may be detecting in our deSFH the small population of low metallicity stars in disk orbits discussed by several authors \citep[e.g.][]{Sestito20, Fernandez-Alvar2021-MPdisk, Nepal2024-oldDisk, Fernandez-Alvar2024-MPTD}

\section{Star formation histories of Milky Way kinematically selected populations}

Obtaining the full evolutionary picture of a complex chemodynamic system like the Milky Way requires the combination of age information with chemistry and kinematics. The most common approach is to select populations with particular chemodynamic properties from a catalogue of stars including stellar ages. In ChronoGal, we proceed in reverse order: we first perform a chemodynamical selection of the stellar populations, which have an associated CMD, and then we derive the deSFH and age-metallicity distribution by fitting that CMD. \citet[][and this conference]{Fernandez-AlvarThinThick} performed a kinematic selection of thin and thick disk stars in the Toomre diagram. The resulting deSFHs are strikingly different, with the thick disk stellar populations being older than 10 Gyr, the age after which the thin disk populations unfold, with similar characteristics as those in the deSFH of the thin-disk-dominated sphere of 100 pc around the Sun, discussed in the previous section and shown in Figure~\ref{deSFH-gcns}. 

\begin{figure}[t]
  \centering
  \includegraphics[scale=.33]{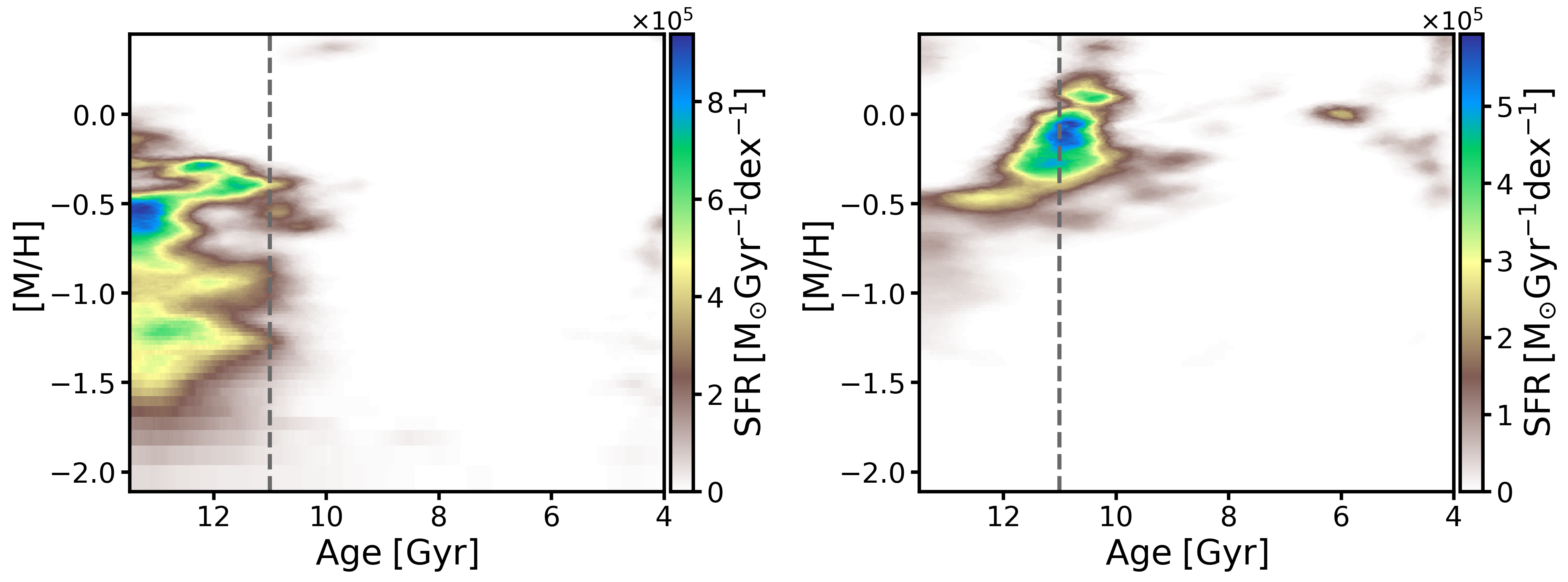}
  \caption{deSFH for the halo and thick disk derived with CMDft.Gaia. Left panel: deSFH for halo stars selected within a cylinder of radius 1 kpc and $200\leq |Z| \leq$ 3500 pc, using tangential velocities calculated from proper motions, as defined in \citet[][]{babusiaux2018}. Right panel: deSFH for a kinematically defined thick disk population selected from a cylinder of radius 250 pc from the Sun and $|Z| \leq$ 500 pc \citep[see][and online proceedings of this conference]{Fernandez-AlvarThinThick}.}
  \label{Halo_ThickDisk}
\end{figure}
Figure \ref{Halo_ThickDisk} shows the deSFH for the thick disk, together with that of the kinematic halo \citep[selected using tangential velocities calculated from proper motions, as defined in][]{babusiaux2018}. The halo CMD selected with this criteria is characterized by two parallel sequences.  The blue sequence has been shown to correspond to an accreted galaxy, Gaia-Sausage-Enceladus (GSE), smaller than the Milky Way progenitor, while the red sequence is mainly composed by in-situ early disk stars heated to halo kinematics by the energy of the merger \citep{belokurov2018_sausage, babusiaux2018, helmi2018, haywood2018, Gallart2019Gaia, diMatteo2020FlatRound, Belokurov2020Splash}. The halo deSFH (left panel) is characterized by a small age range ($\simeq$ 3 Gyr) and a wide metallicity range, extending from [M/H] $\simeq -2$ to almost solar metallicity. The metal-poor portion ([M/H] $\lesssim -0.7$) reflects the stellar populations of GSE. Under the assumption that a dwarf Galaxy stops forming stars when it is accreted into a larger system, the absence of stars younger than $\simeq$ 10.5 Gyr provides a reasonable estimate of the time of GSE accretion \citep[see also][]{Gallart2019Gaia}. The thick disk deSFH, shown in the right panel, is strikingly different. It is characterized by a minor metal-poor population, which more prominent part is a feature with [M/H]$\simeq -0.5$ and age older than 11 Gyr, followed by a major event of star formation with metallicity rapidly increasing up to solar, which average age coincides with the time estimated for the accretion of GSE. This may indicate that the merger of GSE induced a major event of star formation in the early Milky Way disk which produced most of the stars in the thick disk, and enriched the gas up to solar metallicity and even slightly above.

\section{Summary and outlook}
We have shown that CMD-fitting is able to provide a detailed, precise and reliable picture of the age and metallicity distribution of Milky Way stellar populations. The two examples discussed in this contribution are already providing tentative answers to long-lasting questions of Galactic Archaeology, outlined in the introduction. The analysis of datasets that sample the Galaxy more extensively and include other components, such as the bulge, will provide an unprecedentedly precise picture of the Milky Way's evolution.

The current results allow us to provide a basic chronology of the Milky Way evolution, as follows: star formation in a thick disk started earlier than 12 Gyr ago, with a minor amount of stars with [M/H]$\lesssim$-0.5 and a first more prominent $\simeq$ 12 Gyr old population with [M/H]$\simeq$-0.5. Then, around 11 Gyr ago, GSE merged with the Milky Way and triggered a massive star formation event that produced most of the mass in the thick disk and increased its metallicity to solar values. Subsequently, the bulk of the star formation in the thin disk started and continues with a somewhat episodic behaviour up to the present time.

This tentative Milky Way chronology needs to be confirmed and refined using larger and more precise datasets, such as the forthcoming Gaia DR4, expected by the end of 2026. These data will be complemented with those of existing (e.g. APOGEE, GALAH, LAMOST) and new spectroscopic surveys (e.g. WEAVE, 4MOST), which will allow us to select Milky Way populations based on their kinematics and chemistry. Their analysis with CMD-fitting will allow us to secure a solid picture of the Milky Way formation and evolution, as well as of the physical processes involved in the evolution of galaxies in general.

\begin{acknowledgements}
We want to congratulate and thank Beatrice for being such an inspiration in astrophysics, especially for women, and we wish her all the best in her career ahead. EFA and CG acknowledge support from HORIZON TMA MSCA Postdoctoral Fellowships Project TEMPOS, number 101066193, call HORIZON-MSCA-2021-PF-01, by the European Research Executive Agency. EFA, CG, ABQ and AR also acknowledge support from the AEI-MCINN under grants “At the forefront of Galactic Archaeology: evolution of the luminous and dark matter components of the Milky Way and Local Group dwarf galaxies in the {\it Gaia} era” with references PID2020-118778GB-I00/10.13039/501100011033 and PID2023-150319NB-C21/10.13039/501100011033. TRL acknowledges financial support by the research projects AYA2017-84897-P, PID2020-113689GB-I00, and PID2020-114414GB-I00, financed by MCIN/AEI/10.13039/501100011033, the project A-FQM-510-UGR20 financed from FEDER/Junta de Andaluc\'ia-Consejer\'ia de Transformaci\'on Econ\'omica, Industria, Conocimiento y Universidades/Proyecto and by the grants P20-00334 and FQM108, financed by the Junta de Andaluc\'ia (Spain), as well as Juan de la Cierva fellowship (IJC2020-043742-I). SC acknowledges financial support from PRIN-MIUR-22: CHRONOS: adjusting the clock(s) to unveil the CHRONO-chemo-dynamical Structure of the Galaxy” (PI: S. Cassisi) funded by European Union – Next Generation EU, and Theory grant INAF 2023 (PI: S. Cassisi). 
\end{acknowledgements}
\bibliographystyle{aa}
\bibliography{bib_gaia.bib}

\end{document}